\documentclass[12pt]{article}
\pdfoutput=1
\usepackage[top = 2.5 cm, bottom = 2 cm, left = 2 cm, right = 2 cm]{geometry}
\usepackage{authblk, cite, color, amssymb, amsmath}
\usepackage[dvipsnames]{xcolor}

\begin{document}

\title{Binary Neutron Star Mergers with Missing Electromagnetic Counterparts as Manifestations of Mirror World}

\author{{\bf Revaz Beradze$^{1}$}, {\bf Merab Gogberashvili$^{1,2}$} and {\bf Alexander S.~Sakharov}$^{3,4}$}
\affil{\small $^1$ Javakhishvili Tbilisi State University, 3 Chavchavadze Avenue, Tbilisi 0179, Georgia \authorcr
$^2$ Andronikashvili Institute of Physics, 6 Tamarashvili Street, Tbilisi 0177, Georgia \authorcr
$^3$ Physics Department, Manhattan College, Riverdale, New York, NY 10471, USA \authorcr
$^4$ EP Department, CERN, 1211 Geneva 23, Switzerland}
\maketitle

\begin{abstract}

We suggest that the major fraction of binary mergers, which might provide gravitational wave signal detectable by LIGO/Virgo, emerged from the hidden mirror sector. Mirror particles do not interact with an ordinary observer except gravitationally, which is the reason why no electromagnetic signals accompanying gravitational waves from mergers with components composed of mirror matter are expected. Therefore, if the dark matter budget of the universe is mostly contributed by the mirror particles, we predict that only about one binary neutron star (neutron star - black hole) merger out of ten, observable by LIGO/Virgo, in particular at favorable conditions relative to the line of sight, could be accompanied by prompt gamma ray burst and other electromagnetic signals suitable for follow up observations. It seems the list of candidate events recorded by LIGO/Virgo during third observational run supports our predictions.

\vskip 3mm
\noindent
PACS numbers: 95.55.Ym; 04.30.-w; 97.60.Jd; 97.60.Lf
\vskip 1mm
\noindent
Keywords: Multi-messenger Astrophysics; Gravitational Waves; Neutron Stars; Dark matter; Black Holes
\end{abstract}

%%%%%%%%%%%%%%%%%%%%%%%%%%%%%%%%%%%%%%%%%%%%%%%%%%%%%%%%%%%%%%%%%%%%%%%%%%%%%%%

In the first two observational runs (O1 and O2) the Advanced LIGO/Virgo detectors have discovered 10 gravitational wave (GW) signals coming from binary black hole (BBH) mergers as well as one signal generated by a binary neutron star (BNS) coalescence event \cite{signalnew}. Recent analyzes of O1 and O2 data seems to reveal eight additional BBH mergers \cite{Venumadhav:2019tad}. Data acquisition and processing of the third run (O3) of the Advanced LIGO/Virgo detectors is currently under way and a list of candidate events is available online \cite{O3}. Up to now more than 50 new candidate events without electromagnetic counterpart have been detected.

The BNS merger GW170817 detected in O2 \cite{ligo170817} was the first (and up to now only one) multi-messenger event captured in both GWs and electromagnetic radiation registered by {\it Fermi} telescope as a short gamma ray burst (SGRB) \cite{GRB} and by other orbital and ground based facilities as SGRB afterglow, kilonova etc (see for review \cite{em1_170817} and references therein). However, none of the BBH events were accompanied by electromagnetic counterpart \cite{GRBBH, GRBBH2}. That is why, it was suggested that all these BBHs were isolated binaries, formed through common-envelope \cite{isolated}, or via chemically homogeneous evolution \cite{chemically}. Some dynamical processes in dense stellar clusters were also considered to be responsible for the formation of these heavy BBHs \cite{dynamical}. All these mechanisms suggested that BBH could merge at rate
\begin{equation} \label{theor}
\mathcal{R}_{\rm theor}^{\rm BBH} \sim 5-10 ~ {\rm Gpc^{-3}~yr^{-1}}~.
\end{equation}
The range (\ref{theor}) is hardly overlapped from bellow with the range of the rate estimated by LIGO/Virgo results
\begin{equation} \label{LIGO}
\mathcal{R}_{\rm LIGO}^{\rm BBH} \sim 100~ \rm Gpc^{-3} \rm yr^{-1} ~.
\end{equation}
Some models \cite{Belczynski2016, Belczynski2017} give higher mergers rate ($\mathcal{R} \sim 200~ \rm Gpc^{-3} \rm yr^{-1}$) on the price of lower metallicity in composition of progenitor stars of the BBHs components.

Recently, it has been proposed that non-observation of electromagnetic counterparts of BBH mergers is due to the fact that these BBHs emerge from the mirror world as a result of evolution of binary star systems made out of mirror matter \cite{mirror1, mirror2}.

Mirror world restores left-right symmetry of nature, suggesting that each ordinary Standard Model particle has its mirror partner with opposite chirality. The fundamental reason for existence of mirror partners has been first revealed in \cite{LeeYang}. The communication (interaction) between ordinary and mirror worlds occurs only via gravity. Thus, mirror particles are invisible for ordinary observer (and vice versa), while GWs pass through between mirror and ordinary world and hence can be detected. For a review of theoretical foundations of mirror world as well as of its cosmological and astrophysical manifestations see \cite{mirror_okun, mirror_khlopov, mirror}.

In case of existence of mirror partners, the Universe, along with ordinary particles, should contain mirror electrons, nucleons, neutrinos and photons. After recombination, mirror atoms, invisible in terms of ordinary photons while being gravitationally coupled to our matter, constitute a viable candidate for dissipative dark matter, with specific implications for the cosmological evolution, formation of large-scale structure (LSS), structure of galaxies and stars etc. \cite{reheating, Berezhiani:2000gw, str1, str2, str3, leptogenesis, blin_khlopov, Berezhiani:2005vv}. The equal cosmological abundance of mirror and ordinary particles would violate the Big Bang nucleosynthesis (BBN) bounds, since the mirror photons, electrons and neutrinos would double the effective number of neutrinos contributing into the Hubble expansion rate \cite{reheating, Berezhiani:2000gw}. However, the addition to the effective number of neutrinos can be suppressed by factor $(T'/T)^4$ in case if the temperature of the mirror world $T'$ is lower than the temperature $T$ of the ordinary one \cite{Berezhiani:2000gw}. This can be provided by conditions for asymmetric reheating of mirror and ordinary worlds, created after inflationary epoch \cite{reheating, Berezhiani:2000gw}. Thus, if after inflation ordinary and mirror sectors are reheated in a way to have different temperatures and then both systems evolve adiabatically, ratio $T'/T$ will remain nearly invariant in all epochs until today \cite{mirror, Berezhiani:2000gw}. The BBN bounds require that $T'/T < 0.5$ \cite{Berezhiani:2000gw}, whereas a stronger limit $T'/T < 0.3$ occurs from constraints imposed by the cosmological LSS formation and data on cosmic microwave background (CMB) radiation measurements \cite{str1,str2}.

Several times colder mirror world implies that the universe expansion rate at the ordinary BBN epoch ($T\simeq 1$~MeV) is determined by the ordinary matter density itself, so that it would be a vanishing contribution of the mirror world into the ordinary light elements production \cite{mirror, Berezhiani:2000gw}. In contrary, at the epoch of nucleosynthesis in mirror world, the contribution of the ordinary world scales as $(T'/T)^{-4}$ and hence turns to be dramatically high. Thus, the abundances of mirror light elements produced at the expansion rate determined by the ordinary matter should contradict to the density of the mirror matter at the epoch of mirror BBN ($T' \simeq 1$~MeV) \cite{mirror, Berezhiani:2000gw}. As it was shown in \cite{Berezhiani:2000gw}, for $T'/T \lesssim 0.3$, the mirror helium mass fraction can reach $75-80 \%$. In particular, for the temperature ratio ranging from 0.6 to 0.1, the mirror helium abundance $Y_4'$ can vary in the range between 0.5 and 0.8 \cite{mirror, Berezhiani:2000gw}. Thus, as it has been pointed out in \cite{mirror}, a universe with substantial amount of mirror matter should be dominated by mirror helium world with, probably, significant abundances of other mirror heavy elements.

The cosmological relevance of the mirror baryons to dark matter is defined by their contribution $\Omega_{B'}/\Omega_B$ into the density budget of the universe and by the temperatures ratio $T'/T$. The early enough decoupling of mirror photons occurring at $T'/T < 0.3$ makes mirror baryons indistinguishable from the canonical cold dark mater (CDM) in observational tests of LSS and CMB power spectra \cite{Berezhiani:2000gw, str1, str2, str3}. This limit applies independently whether mirror baryons constitute dark matter entirely, or only about 20\% fraction of it, $\Omega_{B'}/\Omega_B \approx 1$ \cite{str2, str3}. Thus, if mirror sector is cold enough, cosmological evolution of the density perturbations of mirror matter is compatible with the observed pattern of LSS and the CMB power spectrum, while its collisional and dissipative nature can have specific observable implications for the evolution and formation of the structures at smaller scales, formation of galaxy halos and stars, etc \cite{mirror, str3}.

Dark matter can even be entirely represented by mirror baryons, i.e.
\begin{equation} \label{omega}
\frac {\Omega_{B'}}{\Omega_B} \approx 5~.
\end{equation}
In this case the normal galaxies with masses lager than $10^9M_{\odot}$ might be formed if $T'/T < 0.2$, while the power of smaller galaxies will be suppressed by Silk damping \cite{Berezhiani:2000gw,str1}.

The main problem of model (\ref{omega}) arise when one considers galactic halos. The point is, that it is unlikely that, having the same microphysics as ordinary matter, the mirror matter could form extended spherical galactic halos but instead should form the disk, as usual matter does. The spherical halos can be possible if mirror stars are formed earlier than ordinary stars, and before the mirror matter collapsed into the disk \cite{dama}. We notice, that a model with ${\Omega_{B'}}/{\Omega_B} \approx 1$ is acceptable from the point of view of galactic structures, without additional assumptions \cite{dama, halo}. In fact, the ordinary and mirror baryon densities in a comparable amount can be provided by co-baryogenesis (or leptogenesis) mechanism via common $B-L$ violating interactions between ordinary and mirror particles \cite{leptogenesis}. In this context, the condition $T'/T < 0.2$ leads to a remarkable prediction $1 \le \Omega_{B'}/\Omega_B \le 5$ \cite{str3}, which can naturally explain so called coincidence problem between amounts of the visible and dark matter in the Universe.

Taking into account the characteristics of mirror world described above, in its most comfortable appearance (\ref{omega}), for the coincidence problem, and adopting the theoretical predictions of the BBH merger rate from different models (\ref{theor}), in \cite{mirror1, mirror2} we estimated that the rate of mirror BBH mergers should be amplified by factor $10-15$ with respect to (\ref{theor}) and reads
\begin{equation}
\mathcal{R}_{\rm mirror}^{\rm BBH} \sim 50-100 ~ {\rm Gpc^{-3}yr^{-1}} ~,
\end{equation}
which is well within interval (\ref{LIGO}) estimated from LIGO/Virgo measurements.

In current paper we want to extend the mirror world scenario \cite{mirror1, mirror2} of emerging of BBH on the neutron stars (NS) and estimate BNS and BH-NS coalescence rates. Mirror neutron stars as a possible sources for the LIGO/Virgo signals has already been discussed in \cite{Addazi}. It was shown that, mirror NS physics, hydrodynamical equations, equation of state and form of gravitational waves, should be similar to the ordinary neutron stars physics in our world.

Here we are interested in BNS and BH-NS that will form and merge within a Hubble time. Stars within mass range from $\simeq 8M_{\odot}$ to $\simeq 50M_{\odot}$ will evolve up to a core collapse supernova explosion. The resulting remnant will be a NS if the mass of progenitor star is below $20M_{\odot}$ otherwise a BH will be formed. Such heavy stars should arrive to the stage of supergiants of size about $30R_{\odot}$, at the end of their life. If the initial separation of the components in BNS exceeds the size of the progenitor supergiants, the pure GWs inspiral time will be several orders of magnitude the age of the Universe. In order to merge within the age of the Universe (Hubble time), initial separation of canonical BNS system must be less than five solar radius, $\lesssim 5 R_{\odot}$, in case the inspiral is dominated by radiation of gravitational waves \cite{form1, form2, Burns:2019byj}. This implies that BNS system is formed through a common-envelope stage, where either both massive stars are not distinct or the primary forms a compact object before being enveloped by the secondary during its supergiant phase. Thus, the inspiral proceeds much faster and leads to tighter initial separation of the two compact objects. Also, NSs and BHs inside globular cluster could gravitate towards its center due to dynamical friction, leading to both a higher likelihood of dynamical capture and an accelerated inspiral aided by three-body interactions with other objects~\cite{globular1}. The two compact objects of BNS and BH-NS systems will lose energy to GWs, causing the objects to inspiral towards one another. The inspiral time scale ranges from $\simeq 85$~Myr to Hubble time. Thus, after long time of inspiraling the compact objects merge releasing GWs signal of great luminosity ($\simeq 10^{53}$~erg/s \cite{gwlumi1}). GWs emitted by a binary system are omnidirectional, however not isotropic \cite{gwincl1, gwincl2}. In particular, it is strongest along the total angular momentum axis of the system (for inclination angles $l=0^{\circ},\ 180^{\circ}$) and weakest along the orbital plane ($l = 90^{\circ}$).

In some seconds after merging the NSs (NS in case of BH-NS merger) get disrupted releasing matter which is supposed to accrete the remnant object powering collimated ultra-relativistic polar jets \cite{Burns:2019byj, BNSemsign1} and mildly relativistic quasi-isotropic outflows \cite{kilo1, kilo2, kilo3} that produce known electromagnetic and even neutrino counterparts. The jet results in a short gamma ray burst (SGRB) which is a luminous flare of highly variable up to MeV energy gamma radiation lasting less than 2 seconds~\cite{DAvanzo:2015kdp}. The emission from collimated ultra-relativistic jet can only be detected by an observer within jet opening angle, $\theta_i$, due to Doppler beaming limiting the visibility region to inverse Lorentz factor $\Gamma^{-1}$ (typically $\Gamma\sim 100$). Being launched, the jet must propagate trough to successfully break-out and move outwards at nearly speed of light $c$. The jet reaches the photometric radius where light can escape for the first time and may release the prompt SGRB emission due to collisions of internal shocks.

After the prompt emission the jet is still speeding away interacting with the surrounding circumburst material \cite{Fong:2015oha}. The bulk Lorentz factor of the interacting jet decreases, the observable angle grows, the jet starts to emit synchrotron radiation across nearly entire electromagnetic spectrum, which can be detected from radio to GeV energies \cite{Fong:2015oha} as SGRB afterglow.

The unbound matter from a merger behaves differently than the bound material that powers the ultra-relativistic jet causing a SGRB. In this case, the neutron rich ejecta containing from $10^{-3}M_{\odot}$ to $0.1M_{\odot}$ moves outwards at a  $\sim$0.1-0.3 c. The ejecta expands and rapidly cools down losing the energy through thermal neutrino emission. Eventually, in $\sim$10-100 ms, it enters relatively slow homologous expansion and the period of synthesis of heavy elements in the so-called r-process. Nuclei freshly synthesized by the r-process are radioactive and decay back to stability. The energy released via $\beta$ decays and fission can power a thermal transient lasting days to weeks, commonly known as a kilonova \cite{kilo1, kilo2, kilo3}. Kilonovae are promising electromagnetic counterparts of BNS and BH-NS mergers because their emission is approximately isotropic (in contrast to beamed SGRBs) and can peak at optical, ultra violet and infrared wavelengths. Brightness, duration, and colors of kilonovae are diagnostics of physical processes taking place in mergers \cite{Burns:2019byj, kilo1, kilo2, kilo3}. When ejection energy ends the kilonova cools down and fades. The quasi isotropic ejecta will continue to move outwards to go, in next few months and years, into the nebula phase. At deceleration radius where the ejecta swept up a comparable amount of material from the surrounding environment, it will transform to a blast wave that releases synchrotron radiation in the radio bands \cite{radioKilo1, radioKilo2, radioKilo3}, analogously described as a kilonova afterglow.

The earliest observable signal from BNS and BH-NS mergers are GWs. They are used in both the detection and characterization of these events as well as in providing localization information for search with other instruments. The latest measurements on the local volumetric rate of BNS mergers have been performed by LIGO/Virgo in O1 and O2 and reads \cite{signalnew}:
\begin{equation} \label{LIGObns}
\mathcal{R}_{\rm LIGO}^{\rm BNS} = 110-3840~ \rm Gpc^{-3} \rm yr^{-1}.
\end{equation}
The rate of BH-NS mergers is only bounded from above
\begin{equation} \label{LIGObhns}
\mathcal{R}_{\rm LIGO} ^{\rm BH-NS} <610~ \rm Gpc^{-3} \rm yr^{-1} ~,
\end{equation}
provided that the masses of BHs in binary systems are equal to $5M_{\odot}$.

As we described above the NSs mergers should be accompanied by electromagnetic signals of various wavelengths generated in the evolution of central remnants. First, we focus on the joint searches for GWs and prompt SGRB emission. The total rate of SGRBs depends on the half-jet opening angle distribution, which, in its turn, can be estimated with the measured jet break in the afterglow light curve \cite{breakAfter1}. Since, the jet breaks are rarely detected the total rate of SGRBs is quite uncertain and reads $1109^{+1432}_{-657} ~ {\rm Gpc^{-3}yr^{-1}}$~\cite{Jin:2017hle}. Moreover, neglecting one particular SGRB with unusually narrow half-jet opening angle one arrives to $162^{+140}_{-83} ~ {\rm Gpc^{-3}yr^{-1}}$ \cite{Jin:2017hle}. Both estimates lie within the range (\ref{LIGObns}) inferred by LIGO/Virgo from observation of GW170817. Because of the solid angle effect only a fraction of successful SGRBs jets can be oriented towards the Earth. Since the emission of GWs is omnidirectional one expects that this fraction of SGRBs would be associated with gravitationally detected BNS or BH-BNS mergers. Recall, that the gravitation emission is stronger when a binary system is more face on. Therefore, the ultra-relativistic jet, which is believed to be aligned with the polar axis of the remnant, should correlate with the strength of the GW signal. The observed inclination angle probability density for gravitationally detected NS mergers is maximized at $l\simeq 30^{\circ}$~\cite{gwincl2}. It was found in \cite{Burns:2019byj} that a comparison of the observed inclination angle distribution for NS mergers detected through GWs \cite{gwincl2} and prompt SGRBs distribution \cite{Fong:2015oha} tells us that roughly 1 in 8 BNS (BH-NS) mergers with detected gravitational signal can spot the Earth within the jet angle. If we restrict ourselves only with the more likely face on systems similar to those like GW170817, where $l < 55^{\circ}$ \cite{ligo170817}, the joint observation rate of GW and SGRB signals can become somewhat higher, namely 1 in about 5. Finally, assuming an improvement in signal to noise ratio in gravitational measurements of the NS systems, which would probably give $l < 27^{\circ}$ for GW170817 \cite{incl170817}, one may expect a sort of 1 to 1 correspondence for systems less than $30^{\circ}$ off from being face on.

Beyond SGRB detection, the joint observations of GWs and electromagnetic counterparts can be performed through finding of SGRB afterglows, kilonovae, and other expected electromagnetic transients from NS mergers. In general, follow-up searches should not expect to recover those events that occur near the Sun \cite{Burns:2019byj}. For the space based searches the constraint spreads out over about $45^{\circ}$ of the Sun, for many narrow-field space-based telescopes like Swift, Hubble, Chandra etc. It is clear that the ground-based limitations are stricter, namely a few hours of RA from the Sun, for a compatible size exclusion zone obtained from the GWs triggering facilities. An exception could be events detectable long enough that the Sun moves across the sky, which typically requires months of detectibility. Usually, the events are getting identified within a week or so. Either case rules out about 15 \% of the sky. Also, one should not rely on recovery of SGRB afterglow and kilonovae in case they occur within several degrees of the galactic plane, because of extinction and high rate of transients at lower energies. Following the claim of \cite{Burns:2019byj}, one can accept that follow-up observations are capable to recover up to 80\% of GW triggers.

We assume that every BNS (BH-NS) system triggered in GWs should produce observable signatures in electromagnetic messengers across wide ranges of energies and time. The particular attention we pay the canonical types of counterparts such as SGRBs, SGRB afterglows and kilonovae. Relying on the discussion above, we accept that it is very likely that at least one type of the counterpart should be recovered in prompt and follow up observations. Since all analyzed canonical signals from NS mergers are brighter when observed from polar position than from equatorial one, especially likely recovery is expected for those BNS (BH-NS) mergers detected not too much off (that means within $\sim 30^{\circ}$) from its face on position with respect to GWs observations.

Data acquisition and processing of O3 is currently under way and a list of candidate events is available online \cite{O3}. By the time of last edition of this article 10 BNS and BH-NS candidate events with probabilities from $61\%$ to $> 99\%$ have been detected, including the compact merger GW190425 \cite{ligo190425} with a total mass $3.4^{+0.3}_{-0.1}M_{\odot}$ being significantly different from the Milky Way population of BNSs that merge within Hubble time and from that of the first detected BNS merger GW170817 \cite{ligo170817}. However, still no one electromagnetic counterpart has been reported yet. Of course, not each of these candidates is well localized because the lack of triggers in one of three detectors. Moreover, some candidates are detected with quite high probability of false alarm. Also, because of the lack of the information about the estimation on inclination angles one cannot be confident that each ultra-relativistic jet putatively producing a SGRB is somehow aligned with the polar axis of the remnant and hence well orientated for recovery of the SGRB or its afterglow. Kilonovae counterparts could also escape the detection as being subluminous in the BNS range of $\simeq 200$~Mpc accessible for current advanced LIGO/Virgo sensitivity or being localized around Sun or the galactic plane. Although, the joint GW-kilonovae detection rate could achieve 21 per year within the BNS horizon of the advance LIGO/Virgo network, as estimated in \cite{Burns:2019byj}.

Ignoring the unknown factors, mentioned above, that potentially disturb the recovery of ether electromagnetic counterpart, we suggest that BNS or BH-NS systems with missing electromagnetic counterparts are formed from stars made out of mirror matter. Following the arguments of \cite{mirror1, mirror2}, we adopt binary systems merger rate formula \cite{bbhnumber}
\begin{equation} \label{R}
\mathcal{R} = \frac{1}{2}~\epsilon \int P(\tau) \dot N_{\rm NS}~d\tau ~,
\end{equation}
with $\epsilon$ being dimensionless efficiency coefficient, $P(\tau)$ the delay time (time elapsed between creation of binary and its merger) and $\dot N_{\rm NS}$ the NS birthrate density which reads
\begin{equation} \label{NNS}
\dot N_{\rm NS} \sim N_{\rm NS}~{\rm SFR(z)} ~.
\end{equation}
Here $N_{\rm NS}$ is number density of NSs, which is usually assumed to be proportional to the total number of stars $N(m)$ in the galaxy of mass $m$,
\begin{equation} \label{N}
N_{\rm NS} \sim N(m)=\frac{m}{\int M\xi(M)~dM}~ ,
\end{equation}
where $\xi(M)$ is a stellar initial mass function that should be integrated in the interval $5~ M_{\odot} < M <25 ~M_{\odot}$ to give a NS as a final product. SFR(z) in equation (\ref{NNS}) is star formation rate which is usually adopted from the best-fit-function of experimental data \cite{SFR}, peaking at $z \sim 1.9$ that corresponds to the lookback time $t \sim 10.3~{\rm Gyr}$. So, the star formation rate, multiplied by the number density of stars that end as a neutron stars, gives the NS birthrate.

We assume that mirror world explains the total amount of the dark matter, i.e. according to (\ref{omega}) the mirror matter contributes into the total density of the universe 5 times more than the ordinary baryons. The peak of SFR is shifted to the earlier time as a consequence of lower temperature of the mirror world. Also, the mirror sector has higher helium abundance, compared to the ordinary world \cite{mirror, Berezhiani:2000gw}. Thus, based on arguments of \cite{mirror1, mirror2}, we can expect about 10 times more BNS and BH-NS binary systems in mirror world with respect to that of ordinary world, so that the merging rates (\ref{LIGObns}) and (\ref{LIGObhns}) are amplified by the same factor $\sim 10$.

The uncertainties of calculated merging rates (\ref{LIGObns})
\footnote{Under the assumption that GW1901425 signal has been generated by BNS coalescence, the local rate of BNS mergers is updated to $\mathcal{R}_{\rm LIGO}^{\rm BNS} = 250-2810~ \rm Gpc^{-3} \rm yr^{-1} $ \cite{ligo190425}.}
and (\ref{LIGObhns}) are large, namely they span over more than one order of magnitude. So almost any sane theoretical estimate can be well reconciled with the results of first two runs of LIGO/Virgo on BNS or BH-NS coalescence rates~\cite{Mapelli:2018wys}. Upcoming results from O3 and future runs will shrink the allowed intervals making models which rely on predictions of rates more falsifiable. However, in our scenario, we can predict with a confidence that most of the detected GW events classified as BNS or BH-NS mergers will not have any electromagnetic counterparts. More precisely, we foresee that roughly only one out of ten BNS mergers will be detected as a multi-messenger event to be seen in both GWs and at least one of the canonical electromagnetic counterparts discussed above. The forecast becomes more rigorous if we restrict ourselves with those BNS (BH-NS) mergers detected not too much off from its face on position with respect to GWs observations. For such favorably inclined systems one can also state that the NS mergers rate derived from the measurements of GWs signals will be $\sim 10$ times higher than that one obtained from SGRB data collected within a particular size of BNS GW sensitivity horizon. Although, in this case, the amount of the observational statistics needed for confirmation of the hypothesis should be substantially increased.

Moreover, the analysis of mirror matter clouds fragmentation into mirror proto-stars reviled that mass of minimal fragments of the mirror matter scales up by factor about 2 with respect to those of ordinary matter, provide that $T'/T \simeq 0.2$ \cite{halo}. Therefore, one might assume that mirror stars are about twice heavier than visible ones. This can explain how it happened that the unusually massive BNS merger GW190425 \cite{ligo190425} formed differently than electromagnetically observed known Galactic BNSs that are expected to merge within a Hubble time \cite{bnsdistr1}.

In case if the mirror world contributes less than total amount of the dark matter, the rate of joint GW electromagnetic observations will increase inversely proportion to the ratio ${\Omega_{B'}}/{\Omega_B}$. Hence the measurements of the joint detection rate potentially can be used as an independent constraint on the cosmology of the mirror world dark matter.

%%%%%%%%%%%%%%%%%%%%%%%%%%%%%%%%%%%%%%%%%%%%%%%%%%%%%%%%%%%%%%%%%%%%%%%%%%%%%%%

\section*{Acknowledgements}

R.B. and M.G. acknowledge the support by Shota Rustaveli National Science Foundation of Georgia (SRNSFG) [DI-18-335/New Theoretical Models for Dark Matter Exploration].

%%%%%%%%%%%%%%%%%%%%%%%%%%%%%%%%%%%%%%%%%%%%%%%%%%%%%%%%%%%%%%%%%%%%%%%%%%%%%%%


\begin{thebibliography}{99}

\bibitem{signalnew}
   B.~P.~Abbott {\it et al.} [LIGO Scientific and Virgo Collaborations],
  ``GWTC-1: A gravitational-wave transient catalog of compact binary mergers observed by LIGO and Virgo during the first and second observing runs,''
  Phys.\ Rev.\ X {\bf 9} (2019) 031040
%  doi: 10.1103/PhysRevX.9.031040
  [arXiv: 1811.12907 [astro-ph.HE]].

\bibitem{Venumadhav:2019tad}
  T.~Venumadhav, B.~Zackay, J.~Roulet, L.~Dai and M.~Zaldarriaga,
  ``New search pipeline for compact binary mergers: Results for binary black holes in the first observing run of advanced LIGO,''
  Phys.\ Rev.\ D {\bf 100} (2019) 023011
%  doi: 10.1103/PhysRevD.100.023011
  [arXiv: 1902.10341 [astro-ph.IM]].

\bibitem{O3}
  GraceDB -- Gravitational-wave candidate event database, LIGO/Virgo O3 public alerts,
  {https://gracedb.ligo.org/superevents/public/O3/}.

\bibitem{ligo170817}
  B.~P.~Abbott {\it et al.} [LIGO Scientific and Virgo Collaborations],
  ``GW170817: Observation of gravitational waves from a binary neutron star inspiral,''
  Phys.\ Rev.\ Lett.\ {\bf 119} (2017) 161101
%  doi: 10.1103/PhysRevLett.119.161101
  [arXiv: 1710.05832 [gr-qc]].

\bibitem{GRB}
  A.~Goldstein {\it et al.},
  ``An ordinary short gamma-ray burst with extraordinary implications: Fermi-GBM detection of GRB 170817A,''
  Astrophys.\ J.\ {\bf 848} (2017) L14
%  doi: 10.3847/2041-8213/aa8f41
  [arXiv: 1710.05446 [astro-ph.HE]].

  \bibitem{em1_170817}
  H.~Ziaeepour,
  ``Binary neutron star (BNS) merger: What we learned from relativistic ejecta of GW/GRB 170817A,''
  MDPI Physics {\bf 1} (2019) 194
%  doi: 10.3390/physics1020018
  [arXiv: 1905.11355 [astro-ph.HE]].

\bibitem{GRBBH}
  E.~Burns {\it et al.} [LIGO Scientific and Virgo Collaborations and Fermi Gamma-ray Burst Monitor Team],
  ``A Fermi gamma-ray burst monitor search for electromagnetic signals coincident with gravitational-wave candidates in advanced LIGO's first observing run,''
  Astrophys.\ J.\ {\bf 871} (2019) 90
%  doi: 10.3847/1538-4357/aaf726
  [arXiv: 1810.02764 [astro-ph.HE]].

\bibitem{GRBBH2}
  J.~L.~Racusin {\it et al.} [Fermi-LAT Collaboration],
  ``Searching the gamma-ray sky for counterparts to gravitational wave sources: Fermi gamma-ray burst monitor and large area telescope observations of LVT151012 and GW151226,''
  Astrophys.\ J.\ {\bf 835} (2017) 82
%  doi: 10.3847/1538-4357/835/1/82
  [arXiv: 1606.04901 [astro-ph.HE]].

\bibitem{isolated}
  N.~Giacobbo and M.~Mapelli,
  ``The progenitors of compact-object binaries: impact of metallicity, common envelope and natal kicks,''
  Mon.\ Not.\ Roy.\ Astron.\ Soc.\ {\bf 480} (2018) 2011
%  doi: 10.1093/mnras/sty1999
  [arXiv: 1806.00001 [astro-ph.HE]].

\bibitem{chemically}
  I.~Mandel and S.~E.~de Mink,
  ``Merging binary black holes formed through chemically homogeneous evolution in short-period stellar binaries,''
  Mon.\ Not.\ Roy.\ Astron.\ Soc.\ {\bf 458} (2016) 2634
%  doi: 10.1093/mnras/stw379
  [arXiv: 1601.00007 [astro-ph.HE]].

\bibitem{dynamical}
  A.~Askar, M.~Szkudlarek, D.~Gondek-Rosińska, M.~Giersz and T.~Bulik,
  ``MOCCA-SURVEY Database - I. Coalescing binary black holes originating from globular clusters,''
  Mon.\ Not.\ Roy.\ Astron.\ Soc.\ {\bf 464} (2017) L36
%  doi: 10.1093/mnrasl/slw177
  [arXiv: 1608.02520 [astro-ph.HE]].

\bibitem{Belczynski2016}
  K.~Belczynski, D.~E.~Holz, T.~Bulik and R.~O'Shaughnessy,
  ``The first gravitational-wave source from the isolated evolution of two 40-100 Msun stars,''
  Nature {\bf 534} (2016) 512
%  doi: 10.1038/nature18322
  [arXiv: 1602.04531 [astro-ph.HE]].

\bibitem{Belczynski2017}
  K.~Belczynski {\it et al.},
  ``The evolutionary roads leading to low effective spins, high black hole masses, and O1/O2 rates of LIGO/Virgo binary black holes,''
  arXiv: 1706.07053 [astro-ph.HE].

\bibitem{mirror1}
  R.~Beradze and M.~Gogberashvili,
  ``LIGO signals from the mirror world,''
  Mon.\ Not.\ Roy.\ Astron.\ Soc.\ {\bf 487} (2019) 650
%  doi: 10.1093/mnras/stz1295
  [arXiv: 1902.05425 [gr-qc]].

\bibitem{mirror2}
  R.~Beradze and M.~Gogberashvili,
  ``Gravitational waves from mirror world,''
  MDPI Physics {\bf 1} (2019) 67
%  doi: 10.3390/physics1010007
  [arXiv: 1905.02787 [gr-qc]].

\bibitem{LeeYang}
  T.~D.~Lee and C.~N.~Yang,
  ``Question of parity conservation in weak interactions,''
  Phys.\ Rev.\ {\bf 104} (1956) 254.
%  doi: 10.1103/PhysRev.104.254

\bibitem{mirror_okun}
  L.~B.~Okun,
  ``Mirror particles and mirror matter: 50 years of speculations and search,''
  Phys.\ Usp.\ {\bf 50} (2007) 380
%  doi: 10.1070/PU2007v050n04ABEH006227
  [hep-ph/0606202].

\bibitem{mirror_khlopov}
  M.~Y.~Khlopov,
  {\it Fundamentals of Cosmic Particle Physics},
  (CISP-Springer, Cambridge 2012).
%  doi:10.1007/978-1-907343-72-8.

\bibitem{mirror}
  Z.~Berezhiani,
  ``Through the looking-glass: Alice's adventures in mirror world,''
  In {\it From Fields to Strings}, vol. 3, eds. M. Shifman, et al. (World Scientific, Singapore 2005) p. 2147-2195
%  doi: 10.1142/9789812775344_0055
  [hep-ph/0508233].

\bibitem{reheating}
  Z.~G.~Berezhiani, A.~D.~Dolgov and R.~N.~Mohapatra,
  ``Asymmetric inflationary reheating and the nature of mirror universe,''
  Phys.\ Lett.\ B {\bf 375} (1996) 26
%  doi: 10.1016/0370-2693(96)00219-5
  [hep-ph/9511221].

\bibitem{Berezhiani:2000gw}
  Z.~Berezhiani, D.~Comelli and F.~L.~Villante,
  ``The early mirror universe: Inflation, baryogenesis, nucleosynthesis and dark matter,''
  Phys.\ Lett.\ B {\bf 503} (2001) 362
%  doi: 10.1016/S0370-2693(01)00217-9
  [hep-ph/0008105].

\bibitem{str1}
  A.~Y.~Ignatiev and R.~R.~Volkas,
  ``Mirror dark matter and large scale structure,''
  Phys.\ Rev.\ D {\bf 68} (2003) 023518
% doi: 10.1103/PhysRevD.68.023518
  [hep-ph/0304260].

\bibitem{str2}
  Z.~Berezhiani, P.~Ciarcelluti, D.~Comelli and F.~L.~Villante,
  ``Structure formation with mirror dark matter: CMB and LSS,''
  Int.\ J.\ Mod.\ Phys.\ D {\bf 14} (2005) 107
% doi: 10.1142/S0218271805005165
  [astro-ph/0312605].

\bibitem{str3}
  Z.~Berezhiani,
  ``Mirror world and its cosmological consequences,''
  Int.\ J.\ Mod.\ Phys.\ A {\bf 19} (2004) 3775
%  doi: 10.1142/S0217751X04020075
  [hep-ph/0312335].

\bibitem{leptogenesis}
  L.~Bento and Z.~Berezhiani,
  ``Leptogenesis via collisions: The lepton number leaking to the hidden sector,''
  Phys.\ Rev.\ Lett.\ {\bf 87} (2001) 231304
%  doi: 10.1103/PhysRevLett.87.231304
  [hep-ph/0107281].

\bibitem{blin_khlopov}
  S.~I.~Blinnikov and M.~Khlopov,
  ``Possible astronomical effects of mirror particles,''
  Sov.\ Astron.\ {\bf 27} (1983) 371
  [Astron.\ Zh.\ {\bf 60} (1983) 632].

\bibitem{Berezhiani:2005vv}
  Z.~Berezhiani, S.~Cassisi, P.~Ciarcelluti and A.~Pietrinferni,
  ``Evolutionary and structural properties of mirror star MACHOs,''
  Astropart.\ Phys.\ {\bf 24} (2006) 495
%  doi: 10.1016/j.astropartphys.2005.10.002
  [astro-ph/0507153].

\bibitem{dama}
  R.~Cerulli, P.~Villar, F.~Cappella, R.~Bernabei, P.~Belli, A.~Incicchitti, A.~Addazi and Z.~Berezhiani,
  ``DAMA annual modulation and mirror dark matter,''
  Eur.\ Phys.\ J.\ C {\bf 77} (2017) 83
%  doi: 10.1140/epjc/s10052-017-4658-3
  [arXiv: 1701.08590 [hep-ex]].

\bibitem{halo}
  J.~S.~Roux and J.~M.~Cline,
  ``Constraining galactic structures of mirror dark matter,''
  arXiv: 2001.11504 [astro-ph.CO].

\bibitem{Addazi}
  A.~Addazi and A.~Marciano,
  ``Testing merging of dark exotic stars from gravitational waves in the multi-messenger approach,''
  Int.\ J.\ Mod.\ Phys.\ A {\bf 33} (2018) 1850167
%  doi: 10.1142/S0217751X18501671
  [arXiv: 1710.08822 [hep-ph]].

\bibitem{form1}
  A.~Sadowski, K.~Belczynski, T.~Bulik, N.~Ivanova, F.~A.~Rasio and R.~W.~O'Shaughnessy,
  ``The total merger rate of compact object binaries in the local universe,''
  Astrophys.\ J.\ {\bf 676} (2008) 1162
%  doi: 10.1086/528932
  [arXiv: 0710.0878 [astro-ph]].

\bibitem{form2}
  J.~A.~Faber and F.~A.~Rasio,
  ``Binary neutron star mergers,''
  Living Rev.\ Rel.\ {\bf 15} (2012) 8
%  doi: 10.12942/lrr-2012-8
  [arXiv: 1204.3858 [gr-qc]].

\bibitem{Burns:2019byj}
  E.~Burns,
  ``Neutron star mergers and how to study them,''
  arXiv: 1909.06085 [astro-ph.HE].

\bibitem{globular1}
  K.~Belczynski, V.~Kalogera and T.~Bulik,
  ``A Comprehensive study of binary compact objects as gravitational wave sources: Evolutionary channels, rates, and physical properties,''
  Astrophys.\ J.\ {\bf 572} (2001) 407
%  doi: 10.1086/340304
  [astro-ph/0111452].

\bibitem{gwlumi1}
  F.~Zappa, S.~Bernuzzi, D.~Radice, A.~Perego and T.~Dietrich,
  ``Gravitational-wave luminosity of binary neutron stars mergers,''
  Phys.\ Rev.\ Lett.\ {\bf 120} (2018) 111101
%  doi: 10.1103/PhysRevLett.120.111101
  [arXiv: 1712.04267 [gr-qc]].

\bibitem{gwincl1}
  B.~F.~Schutz,
  ``Determining the Hubble constant from gravitational wave observations,''
  Nature {\bf 323} (1986) 310.
%  doi:10.1038/323310a0

\bibitem{gwincl2}
  B.~F.~Schutz,
  ``Networks of gravitational wave detectors and three figures of merit,''
  Class.\ Quant.\ Grav.\ {\bf 28} (2011) 125023
%  doi: 10.1088/0264-9381/28/12/125023
  [arXiv: 1102.5421 [astro-ph.IM]].

\bibitem{BNSemsign1}
  R.~Fern\'{a}ndez and B.~D.~Metzger,
  ``Electromagnetic signatures of neutron star mergers in the advanced LIGO Era,''
  Ann.\ Rev.\ Nucl.\ Part.\ Sci.\ {\bf 66} (2016) 23
%  doi: 10.1146/annurev-nucl-102115-044819
  [arXiv: 1512.05435 [astro-ph.HE]].

\bibitem{kilo1}
  B.~D.~Metzger and R.~Fernández,
  ``Red or blue? A potential kilonova imprint of the delay until black hole formation following a neutron star merger,''
  Mon.\ Not.\ Roy.\ Astron.\ Soc.\ {\bf 441} (2014) 3444
%  doi: 10.1093/mnras/stu802
  [arXiv: 1402.4803 [astro-ph.HE]].

\bibitem{kilo2}
  B.~D.~Metzger,
  ``Kilonovae,''
  Living Rev.\ Rel.\ {\bf 23} (2020) 1
%  doi:10.1007/s41114-019-0024-0
  [arXiv: 1910.01617 [astro-ph.HE]].

\bibitem{kilo3}
  M.~Tanaka,
  ``Kilonova/macronova emission from compact binary mergers,''
  Adv.\ Astron.\ {\bf 2016} (2016) 6341974
%  doi: 10.1155/2016/6341974
  [arXiv: 1605.07235 [astro-ph.HE]].

\bibitem{DAvanzo:2015kdp}
  P.~D'Avanzo,
  ``Short gamma-ray bursts: A review,''
  JHEAp {\bf 7} (2015) 73.
%  doi: 10.1016/j.jheap.2015.07.002

\bibitem{Fong:2015oha}
  W.~f.~Fong, E.~Berger, R.~Margutti and B.~A.~Zauderer,
  ``A decade of short-duration gamma-ray burst broadband afterglows: Energetics, circumburst densities, and jet opening angles,''
  Astrophys.\ J.\ {\bf 815} (2015) 102
%  doi: 10.1088/0004-637X/815/2/102
  [arXiv: 1509.02922 [astro-ph.HE]].

\bibitem{radioKilo1}
  E.~Nakar and T.~Piran,
  ``Radio remnants of compact binary mergers - the electromagnetic signal that will follow the gravitational waves,''
  Nature {\bf 478} (2011) 82
%  doi: 10.1038/nature10365
  [arXiv: 1102.1020 [astro-ph.HE]].

\bibitem{radioKilo2}
  T.~Piran, E.~Nakar and S.~Rosswog,
  ``The electromagnetic signals of compact binary mergers,''
  Mon.\ Not.\ Roy.\ Astron.\ Soc.\ {\bf 430} (2013) 2121
%  doi: 10.1093/mnras/stt037
  [arXiv: 1204.6242 [astro-ph.HE]].

\bibitem{radioKilo3}
   K.~Hotokezaka and T.~Piran,
  ``Mass ejection from neutron star mergers: Different components and expected radio signals,''
  Mon.\ Not.\ Roy.\ Astron.\ Soc.\ {\bf 450} (2015) 1430
%  doi: 10.1093/mnras/stv620
  [arXiv: 1501.01986 [astro-ph.HE]].

\bibitem{breakAfter1}
  R.~Sari, T.~Piran and J.~Halpern,
  ``Jets in GRBs,''
  Astrophys.\ J.\ {\bf 519} (1999) L17
%  doi: 10.1086/312109
  [astro-ph/9903339].

\bibitem{Jin:2017hle}
  Z.~P.~Jin {\it et al.},
  ``Short GRBs: opening angles, local neutron star merger rate and off-axis events for GRB/GW association,''
  Astrophys.\ J.\ {\bf 857} (2018) 128
%  doi: 10.3847/1538-4357/aab76d
  [arXiv: 1708.07008 [astro-ph.HE]].

\bibitem{incl170817}
  I.~Mandel,
  ``The orbit of GW170817 was inclined by less than 28$^{\circ}$ to the line of sight,''
  Astrophys.\ J.\ {\bf 853} (2018) L12
%  doi: 10.3847/2041-8213/aaa6c1
  [arXiv: 1712.03958 [astro-ph.HE]].

\bibitem{ligo190425}
  B.~P.~Abbott {\it et al.} [LIGO Scientific and Virgo Collaborations],
  ``GW190425: Observation of a compact binary coalescence with total mass $\sim 3.4 M_{\odot}$,''
  arXiv: 2001.01761 [astro-ph.HE].

\bibitem{bbhnumber}
  O.~D.~Elbert, J.~S.~Bullock and M.~Kaplinghat,
  ``Counting black holes: The cosmic stellar remnant population and implications for LIGO,''
  Mon.\ Not.\ Roy.\ Astron.\ Soc.\ {\bf 473} (2018) 1186
%  doi: 10.1093/mnras/stx1959
  [arXiv: 1703.02551 [astro-ph.GA]].

\bibitem{SFR}
  P.~Madau and M.~Dickinson,
  ``Cosmic star formation history,''
  Ann.\ Rev.\ Astron.\ Astrophys.\ {\bf 52} (2014) 415
%  doi: 10.1146/annurev-astro-081811-125615
  [arXiv: 1403.0007 [astro-ph.CO]].


\bibitem{Mapelli:2018wys}
  M.~Mapelli and N.~Giacobbo,
  ``The cosmic merger rate of neutron stars and black holes,''
  Mon.\ Not.\ Roy.\ Astron.\ Soc.\ {\bf 479} (2018) 4391
%  doi: 10.1093/mnras/sty1613
  [arXiv: 1806.04866 [astro-ph.HE]].

\bibitem{bnsdistr1}
  N.~Farrow, X.~J.~Zhu and E.~Thrane,
  ``The mass distribution of Galactic double neutron stars,''
  Astrophys.\ J.\ {\bf 876} (2019) 18
%  doi: 10.3847/1538-4357/ab12e3
  [arXiv: 1902.03300 [astro-ph.HE]].

\end{thebibliography}
\end{document}